\documentclass[12pt,preprint]{aastex}
\begin{document}
\title{Exploring cloudy gas accretion as a source of interstellar 
turbulence in the outskirts of disks}
\author{A.~Santill\'an\altaffilmark{1}, 
F.J. S\'{a}nchez-Salcedo\altaffilmark{2} and J. Franco\altaffilmark{2}} 
\altaffiltext{1}{C\'omputo Aplicado-DGSCA, Universidad Nacional
Aut\'onoma de M\'exico, Ciudad Universitaria, 04510 Mexico City, Mexico; 
alfredo@astroscu.unam.mx.}
\altaffiltext{2}{Instituto de Astronom\'\i a, Universidad Nacional Aut\'onoma
de M\'exico, Ciudad Universitaria, 
04510 Mexico City, Mexico; jsanchez@astroscu.unam.mx, 
pepe@astroscu.unam.mx.}

\begin{abstract}
High--resolution 2D--MHD numerical simulations have been carried out
to investigate the effects of continuing infall of clumpy gas
in extended H\,{\sc i} galactic disks. Given a certain accretion
rate, the response of the disk depends on its surface gas density and
temperature. For Galactic conditions at a galactocentric
distance of $\sim 20$ kpc, and for mass accretion rates
consistent with current empirical and theoretical determinations in
the Milky Way, the rain of compact high velocity clouds onto the disk
can maintain transonic turbulent motions in the warm phase  
($\sim 2500$ K) of H\,{\sc i}. Hence,
the H\,{\sc i} line width is expected to be $\sim 6.5$ km s$^{-1}$
for a gas layer at $2500$ K, if infall
were the only mechanism of driving turbulence.
Some statistical properties of the resulting forcing
flow are shown in this Letter.
The radial dependence of the gas velocity dispersion is also discussed.

\end{abstract}
\keywords{galaxies: intergalactic medium --- galaxies: ISM --- hydrodynamics
--- ISM: kinematics and dynamics --- ISM: structure --- turbulence}
\section{Introduction}
H\,{\sc i} line widths, $\sigma_{W}$,
 are observed to vary from $\sim 12$ to $15$ km s$^{-1}$
in the central parts to $6$-$8$ km s$^{-1}$ in the outer parts.
Beyond the optical disk, the maps of $\sigma_{W}$
display a patchy distribution with
values populating the interval from $5$ to $9$ km s$^{-1}$,
independent of galactocentric distance.  
The median value is fairly constant along the extended outer parts, 
with about the same universal value $7\pm 1$ km s$^{-1}$ for all the 
observed galaxies 
(\citealt{lew84,dib06}, and references therein).
By fitting two-component Gaussians to the H\,{\sc i}
profiles in NGC 6822, the second-moments for the cold and warm 
phases of the H\,{\sc i} were found to be $\sigma_{W}\approx 5$ km s$^{-1}$ 
and $\sigma_{W}\approx 8$ km s$^{-1}$, respectively \citep{blo06}.

The observed $\sigma_{W}$ are likely to represent turbulent small-scale
motions: 
the existence of a cold phase in the outer disks, 
the observed level of star formation 
and the double exponential radial profile of the star formation rate 
have been also interpreted as a consequence of the turbulence gas compressions
in the outer disks (e.g., \citealt{fer98, elm06}). 

\citet{dib06} found that if turbulence is driven by supernova, 
the velocity dispersion of the H\,{\sc i} gas in the quiescent regime is
$\sim 3$ km s$^{-1}$, a factor of $\sim 2$ smaller 
than the observed values. Thus, 
either there are other physical processes driving turbulence or the
supernova feedback efficiency has been underestimated.
There exist several physical mechanisms that could drive turbulence
even in the absence of star formation: hydrodynamic or
magnetohydrodynamic instabilities, frequent
minor mergers of small satellite clumps, 
ram pressure or infalling gas clouds. Amongst the
hydrodynamical instabilities, the thermal instability by itself
cannot sustain turbulence \citep{san01, gaz01,
bra07}. The magnetorotational instability, on the
other hand, can only account for an amount of turbulent motions $\sim 4.5$
km s$^{-1}$ for the three components, if thermal broadening is
not subtracted, 
and might be completely suppressed by stellar
feedback (Dib et al.~2006 and references therein). 

Our main goal is to assess how
much of the velocity dispersion observed in the ISM 
can be due to the impact of high velocity clouds (HVCs) 
and intermediate velocity clouds (IVCs). 
Ample evidence for the presence of continuing gaseous infall 
to the Galactic disk has been compiled in \citet{bec03}.
All the studies firmly suggest that the inflow in the Galactic
disk has been constant, with an accretion rate of $\sim 1$ M$_{\odot}$ 
yr$^{-1}$, or has even increased, during its lifetime.
H\,{\sc i} observations have revealed accretion of both diffuse
and discrete structures from the extended environment in M31 and
M33 \citep{thi04, bra04}.
\citet{put06} infers that the present total mass in condensed infalling
clouds around the Milky Way is $\sim 6\times 10^{8}$ M$_{\odot}$ 
if they are all at
distances $<60$ kpc. HVCs may be a repository for large amounts of gas
if clouds fall into the disk rapidly after they are formed 
\citep{mal03,mal04,put06}.
In the next sections, we consider the turbulent H\,{\sc i}
dynamics driven by the infall of clouds. Preliminary results
were reported in S\'anchez-Salcedo et al.~(2007).

\section{The model}
The ideal MHD equations are integrated using the ZEUS
code (Stone \& Norman 1992a,b). 
A local Cartesian frame of reference with $x$ and $z$ 
corresponding to the horizontal and vertical directions, respectively,
was adopted to simulate a small patch of our Galaxy at a distance
$R_{0}$ from the Galactic center.
The galactic symmetry plane, $z=0$, is placed in the middle
of the computational domain. 
The simulation domain is a square of size 
$L\times L$ with $1024^{2}$ zones.
The ambient medium that is interacting with the HVCs is initially in  
plane-parallel magnetohydrostatic equilibrium with scaleheight
$h_{0}$ in an external gravitational potential.
Here the scaleheight is defined
as half width at half maximum of the vertical volume density of the
gas layer.  The ambient gas is initially isothermal in space
with {\it thermal} sound speed $c_{s}$ and its evolution is 
nearly isothermal with specific heat ratio $\gamma=1.01$. 
Gas self-gravity was disregarded.
The magnetic field has only one component along the $y$--axis 
and is initially stratified in the $z$--direction, $B_{\rm y}$($z$). 
Thus, there is no magnetic tension, only magnetic compression.
Two magnetic configurations are explored. 
The first one is the high-latitude extension of the
Galactic thick disk of Boulares \& Cox (1990, BC model hereafter) 
following \citet{san99}.
The vertical gravitational force and the magnetic field
configuration were scaled to the outer Galaxy,
neglecting edge effects. The second set of models assumes 
that $B_{y}^{2}$ falls off
with the gas density (Spitzer case) in the initial equilibrium configuration, 
so that $\alpha$, defined as the ratio
between the magnetic and thermal pressure, is constant. 
In all the simulations, the size of the domain
is $L>10 h_{0}$ in order to avoid
spurious boundary effects.  We apply periodic boundary 
conditions in the $x$--axis, and open boundary conditions on the upper
and lower faces of the computational domain.

The too low thermal gas pressure at the outer parts of extended 
galactic disks will permit only the warm (several times $10^{3}$ K)
phase of H\,{\sc i} to exist (e.g., \citealt{elm94}).
\citet{hei01} provides observational evidence that 
about half of the mass of the diffuse gas in the Galaxy may have
temperatures from a few hundred to a few thousand kelvin.
In this work, we will explore $c_{s}$ values in the range 
$4.5$--$8$ km s$^{-1}$.

The ambient gas is subject to a continuous rain of HVCs.
In most simulations, if not specified otherwise, 
all the clouds with initial radius $R_{cl}$ and uniform internal density 
$n_{cl}$, are injected at $z=z_{cl}$
with a vertical velocity $v_{cl}$, and are placed randomly in $x$.  
We found that the statistical properties of the flow in the disk
are similar when the injection occurs along both caps $z=\pm z_{cl}$.

The remaining five dimensionless parameters that characterize
our simulation models are $\alpha(z)$, $R_{cl}/h_{0}$, $v_{cl}/c_{s}$, 
$n_{cl}/n_{0}$, and $\lambda_{\rm acc}$,
where $n_{0}$ is the 
midplane volume density at $t=0$ and $\lambda_{\rm acc}$ is the
fraction of accreted mass in one sound crossing time, 
$\lambda_{\rm acc}\equiv \left(h_{0}/c_{s}\right)(\dot{\Sigma}/\Sigma_{i})$ 
with $\Sigma_{i}$ is the initial column density of the disk.
Note that once $c_{s}$ is fixed, 
the scaleheight $h_{0}$ is determined by the depth of the external 
gravitational potential. 
The collisions of infalling clouds onto the gaseous disk
will take the gas out of hydrostatic
equilibrium. Consequently, the scaleheight $h$ of the H\,{\sc i}
layer may evolve in time.

\section{Results}
The disk gas is subject to a stocastic driving force caused
by the MHD pull of randomly-injected clouds,
plus the restoring gravitational force which tends to push back the
gas to the midplane.
The level of substructure in the resulting fluid depends on the structural
parameters of the clouds and injection velocity.
We focus first on describing the evolution of our fiducial BC model,
labelled as run BCa,
that is intended to represent conditions at a galactocentric distance
$R_{0}=20$ kpc.
In this model, the parameters of the disk are 
$c_{s}=8$ km s$^{-1}$, $\Sigma\simeq 1.2$ M$_{\odot}$ pc$^{-2}$ and $h_{0}=220$ pc. 
For the clouds we adopted $v_{cl}=100$ km s$^{-1}$, $z_{cl}=2$ kpc, 
$n_{cl}=0.1$ cm$^{-3}$ and $R_{cl}=50$ pc (see S\'anchez-Salcedo et al.~2007
for a discussion of this choice).  The injection rate of clouds in the
whole box domain is $300$ per Gyr, corresponding to an accretion rate 
defined as $4\pi R_{0}^{2}\dot{\Sigma}$ of $0.6$ M$_{\odot}$ yr$^{-1}$. 
While our simulations are 2D, this situation is energetically equivalent
to an accretion rate of $750$ clouds per Gyr and per kpc$^{2}$
of galactic disk. Since the kinetic energy of a cloud is $\sim 10^{50}$ erg,
the energy input rate from impacting clouds is $7.5\times 10^{52}$
erg Gyr$^{-1}$ kpc$^{-2}$, which is equivalent to $75\epsilon^{-1}$ 
supernova explosions per Gyr and kpc$^{2}$ of disk, each
with mechanical energy $\epsilon\times 10^{51}$ erg, where $\epsilon \sim 0.1$ 
is the efficiency factor. At the end of the simulation ($2$ Gyr), the
total mass in our grid increased by a factor $1.45$.

\begin{figure}
\plotone{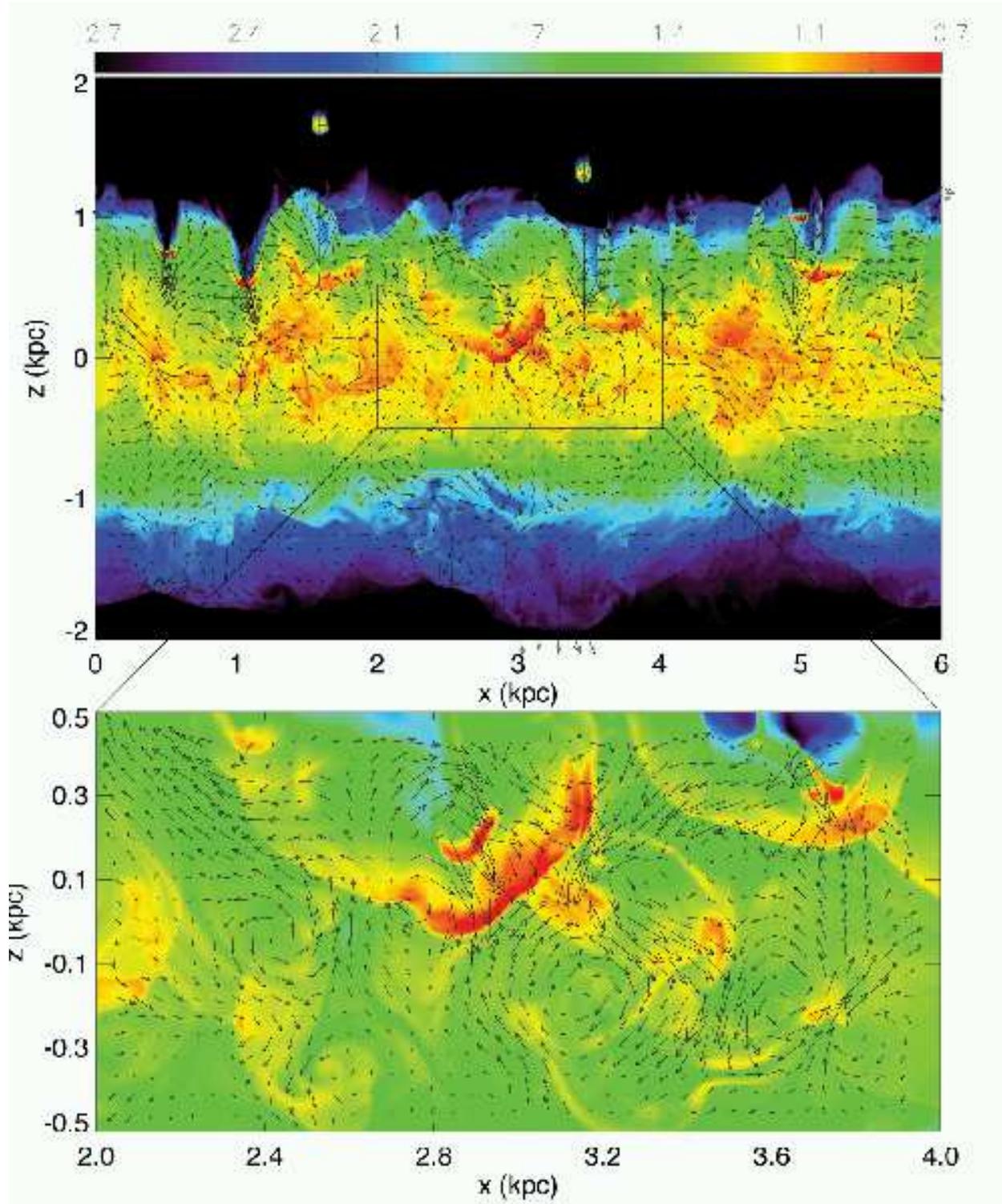}
\caption{Density ({\it color logarithmic scale}) and velocity
field ($arrows$) at $t = 1.7$ Gyr for our run BCa.
A general view of the disk between $-2$ kpc $<z< 2$ kpc
(note that the $z$-axis has been cut-off) is shown ($top$).
A magnification zoom right
into the marked region is displayed ($bottom$).
Notice the presence of vortical motions and the dilution of the 
infalling clouds. }
\label{fig:density-velocity}
\end{figure}

\begin{figure}
\plotone{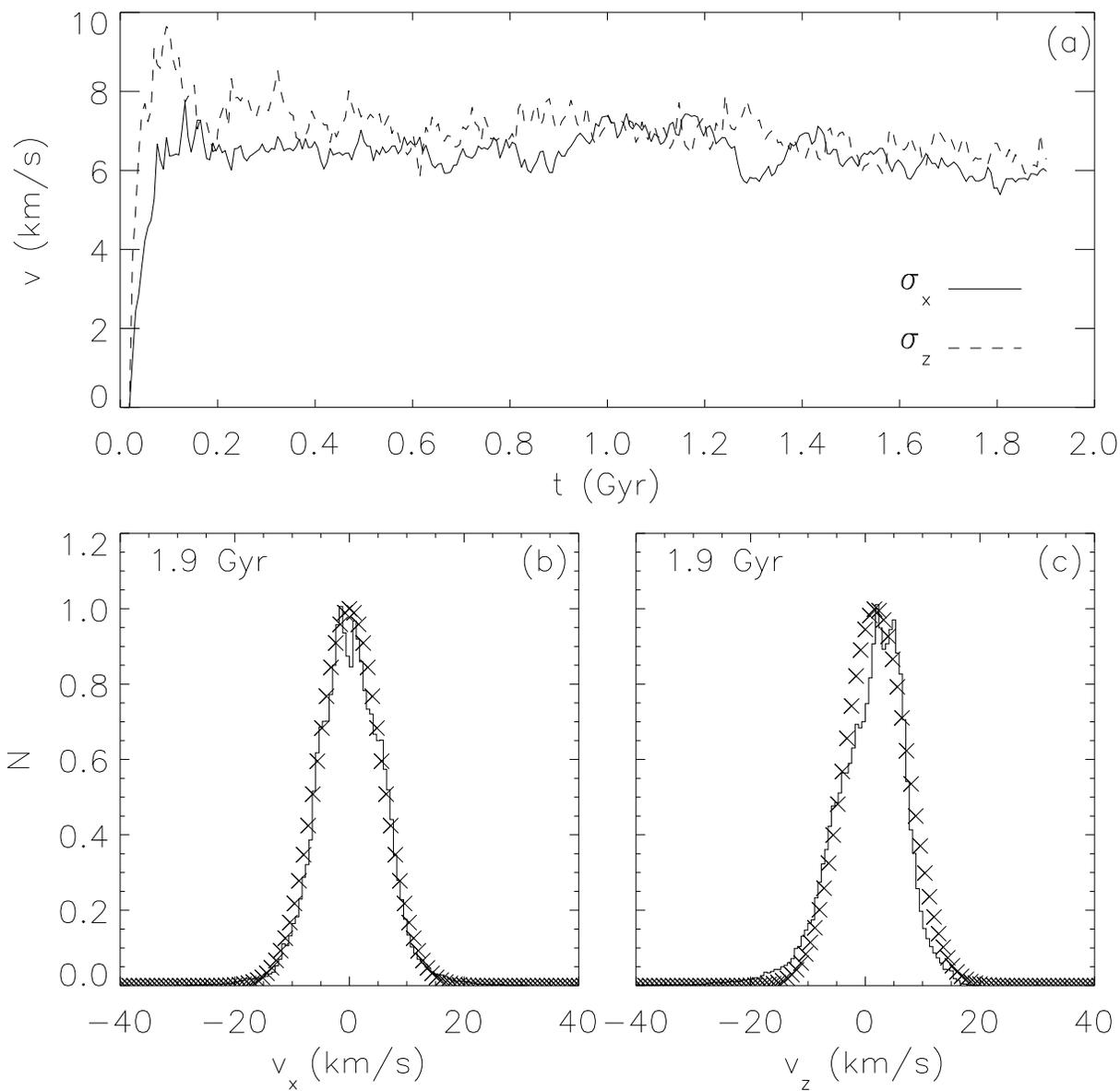}
\caption{Horizontal ({\it solid line}) and vertical
({\it dashed line}) velocity dispersions, with thermal broadening
subtracted, for our fiducial case (run BCa), 
as a function of time (panel {\it a}). 
Panels {\it b} and {\it c} show the mass-weighted profile at $t=1.9$ Gyr.
The profiles sample those gas elements within one scaleheight of the disk.
Overplotted are the Gaussian fits (crosses).         }
\label{fig:rms-velocity}
\end{figure}

Clouds interact with the ambient medium and dilute 
(see Fig.\ref{fig:density-velocity}). At a height
of $500$ pc they lose their identity and mix up with the turbulent disk. 
Hence the number of `active' clouds is $300$ Gyr$^{-1}\times (1.5\,{\rm kpc}/
100\,{\rm km \, \,s}^{-1}) \sim 4$--$5$.
The infalling clumpy flow perturbs the disk through MHD waves that
interact in a complicated fashion, producing a network of plumes and
shells (see Fig.~\ref{fig:density-velocity}). As a consequence, 
the 1D rms velocity of the gas, $\sigma$,
increases in time and saturates after $100$ Myr,
reaching values that oscillate between $6$ and $7.5$ km s$^{-1}$
around the mean value $\bar{\sigma}$
(see Fig.~\ref{fig:rms-velocity}). The difference between 
$\bar{\sigma}_{x}$ and $\bar{\sigma}_{z}$ is only $\sim 0.5$ km s$^{-1}$. 
The velocity dispersion in each
direction was obtained by fitting the mass-weighted velocity profile
for cells within $|z|<h_{f}$, with $h_{f}$ the asymptotic value
of the scaleheight. Examples of mass-weighted velocity profiles are given in
Figure \ref{fig:rms-velocity}b,c.
Remind that $\sigma_{W}=(\sigma^{2}+c_{s}^{2})^{1/2}$ for isothermal gas,
thus, $\sigma_{W}\sim 10$ km s$^{-1}$.
For $v_{cl}=50$ km s$^{-1}$, $\sigma_{x}$ and $\sigma_{z}$ reach
similar values as in the case $v_{cl}=100$ km s$^{-1}$.
These results suggest that continuous accretion of a clumpy gas may
contribute significantly to the random motions observed.

\begin{figure}
\plotone{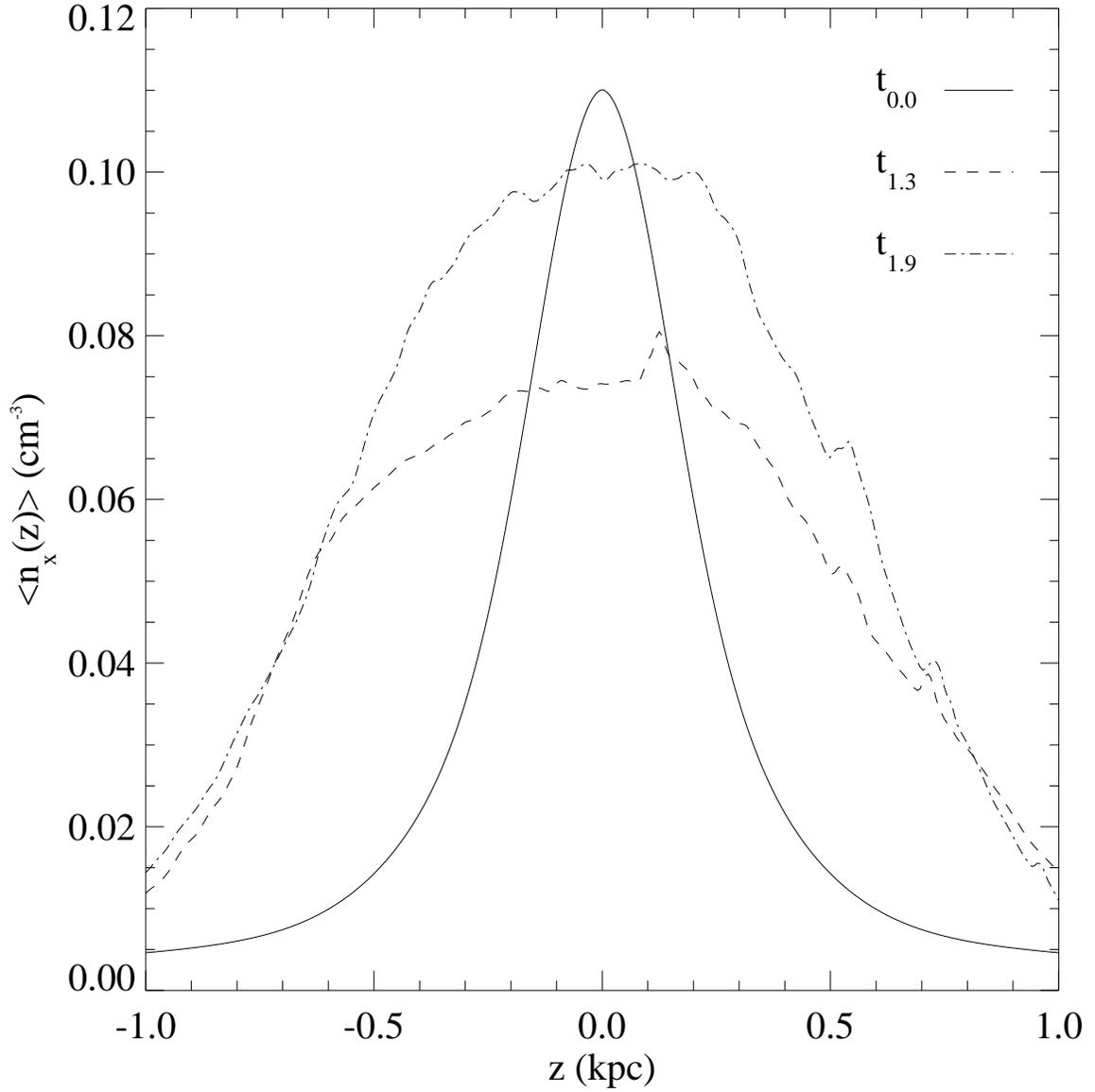}
\caption{The vertical density profile averaged on horizontal cuts is shown
at three selected times, $t = 0, 1.27$ and $1.9$ Gyr, for 
run BCa.  The corresponding
scale heights are $220$, $660$ and $640$ pc, respectively.}
\label{fig:z-density}
\end{figure}

The vertical density profile averaged on horizontal cuts is shown 
in Fig.~\ref{fig:z-density}, 
Note that even though the injection of clouds occurs only through the
upper cap $z=z_{cl}$, the density profile is quite
symmetric and the net vertical displacement is relatively small. 
The scaleheight is $\sim 640$ pc at $\sim 2$ Gyr, similar to the observed 
value at $R=20$ kpc.

The injected energy is mostly dissipated in radiative shell compressions and
behind shock fronts, being the dissipation timescale for turbulent
kinetic energy $\simeq 8$ Myr.
The energy spectrum contains information about how the kinetic energy
is distributed at different scales and about the nature
of the driving mechanisms of the random motions in the 
compressible ISM. Due to the stratification and the anisotropy of
the random motions, we have calculated the
power spectrum along horizontal cuts for each component of
the velocity separately.  The power spectrum
follows a power-law with different exponent for each
component $E_{i}(k)\sim k^{-\alpha_{i}}$, with $\alpha_{x}=3$
and $\alpha_{z}=4$.

\citet{vaz94} and \citet{pad97} reported
lognormal density probability distribution functions (PDF) for
two- and three-dimensional isothermal turbulent flows. 
We have checked that, in fact, our PDF is approximately lognormal
with excesses at large and low densities probably because in our case
the turbulence is not isotropic and,  in addition, there is some probability
of finding density substructure that has not a turbulent origin, it being
the remnants of infalling clumps. 

When the clouds are randomly injected along both caps, 
the mass enhancement in the central disk ($|z|<600$ pc) is slightly
larger (about $10\%$)
than when the injection only proceeds along one side, because
the symmetry in the injection leads to a more efficient balance of momentum.
However, $\sigma(t)$ is quantitatively similar and thus there is no 
need to show it again. 

The elasticity of the disk to collisions depends on the speed of
sound and on the magnetic field strength.
In order to see the sensitivity of $\sigma$ on the 
adopted magnetic configuration, we started
with an $\alpha=1$ Spitzer model and the same $c_{s}$ and 
total pressure at the midplane as in run BCa. 
$\bar{\sigma}_{x}$ and $\bar{\sigma}_{z}$ were found to be
$5$ and $6.2$ km s$^{-1}$, respectively.

It turns out very difficult to infer from basic principles how $\bar{\sigma}$
scales with $c_{s}$ or with any other parameter (S\'anchez-Salcedo
et al.~2007). In order to gain some insight, 
Spitzer configurations with $\alpha=1$ but different $c_{s}$, 
between $5$ and $12$ km s$^{-1}$ (spaced by $1$ km s$^{-1}$),
were considered.  If the total pressure at the 
midplane $z=0$ is identical in all the models
and equal to the value in the standard model\footnote{The fact 
that the total pressure at $z=0$ 
is the same in these models implies that the surface gas density
increases for low values of $c_{s}$.}, we find that
$\bar{\sigma}$ scales linearly with $c_{s}$; a good fit to the data is 
$\bar{\sigma}_{z}=0.55 c_{s}+1.16$ (thermal broadening has been
subtracted) when $v_{cl}=50$ km s$^{-1}$. 
If, instead of the total pressure, the
surface gas density is taken the same as in the case BCa in all the
experiments, then
$\bar{\sigma}_{z}\propto \sqrt{c_{s}}$. For instance, using again
$v_{cl}=50$ km s$^{-1}$, 
$\bar{\sigma}_{z}$ varies from $7.4$ km s$^{-1}$ to $5.0$ km s$^{-1}$ 
when the sound speed
drops by a factor of $2$ (from $12$ km s$^{-1}$ to $6$ km s$^{-1}$).

In order to simulate various galactic conditions at different galactocentric
distances, runs have been carried out varying $\lambda_{\rm acc}$
and $R_{cl}/h_{0}$. Assuming that $R_{cl}$ does not depend strongly on radius,
$R_{cl}/h_{0}$ should decrease with $R$ because the observed $h$ increases
with $R$, but it is uncertain as $R_{cl}$ could also increase with
$R$ as well. The radial variation of $\lambda_{\rm acc}$ is also uncertain
because the accreting mass flux $\dot{\Sigma}(R)$ and the 
temperature structure $c_{s}(R)$ should be known.
As a departure assumption consider that accretion of an intergalactic
plane-parallel flow at infinity occurs ballistically 
$\dot{\Sigma}\propto R^{-1/2}$ (L\'opez-Corredoira et al.~2002).
For our Galaxy, $\Sigma$ decreases by a factor of $\sim 10$ between $15$ and
$25$ kpc, whereas $h$ increases by a factor of $4$.  
Thus, $\lambda_{\rm acc}$ varies by a factor of $\sim 30$ assuming
that $c_{s}$ is constant with $R$.
In order to bracket conditions at these galactocentric distances, 
$\lambda_{\rm acc}$ has been varied by a factor of $25$ for run BCb 
as compared to run BCc, whereas $R_{cl}/h_{0}$ varies by a factor of $4$
(see Table \ref{table:T1}). In runs BCb and Sp-b,
$\sigma_{z}$ reaches a maximum and decreases
afterwards partly because the accreted mass in the disk
is no longer negligible in the length of the run. Since in these
experiments $v_{z}$
deviates significantly from a Gaussian distribution,
whereas $v_{x}$ is still well described by a Gaussian,
only the values of $\bar{\sigma}_{x}/c_{s}$ 
are reported in Table \ref{table:T1}.
According to run BCc, which represents Galactic conditions at $R\sim 15$ kpc,
we find that for $c_{s}=4.5$ km s$^{-1}$, 
$\bar{\sigma}_{x}\approx 3.4$ km s$^{-1}$ and $\sigma_{W}=5.6$
km s$^{-1}$ if thermal broadening is taken into account. At $c_{s}=8$
km s$^{-1}$, $\bar{\sigma}_{x}\approx 4.5$ km s$^{-1}$ (thus $\sigma_{W}=9$
km s$^{-1}$).  At the outer far disk ($\sim 25$ kpc),   
$\bar{\sigma}_{x}$ is found to be $7.5$--$10$ km s$^{-1}$ for
$c_{s}=4.5$ and $8$ km s$^{-1}$, respectively 
($\sigma_{W}\sim 9$--$13$ km s$^{-1}$). 
To derive these values, we rescaled the results of run BCb appropriately
using our previous result that $\bar{\sigma}_{x}\propto \sqrt{c_{s}}$.
Slightly smaller $\sigma$'s are derived for Spitzer models (see Table
\ref{table:T1}).
Line widths of $\sim 10$ km s$^{-1}$ as predicted at the edge of the disk seem
rather high, but note that 
empirical determinations are scarce and very uncertain 
at low surface gas densities ($\Sigma< 1$ M$_{\odot}$ pc$^{-2}$) 
(e.g., \citealt{sho84}). 

If accretion occurs through rare mergers of very massive clouds, 
it would be hard to see how such a localized events
could produce the observed uniform level of turbulence.  
Nevertheless, experiments with input rates of new clouds $25$ times smaller 
than in our fiducial simulation (i.e., $12$ per Gyr in our box domain) 
but cloud radii $5$ times larger ($250$ pc in physical units), so 
that the mass accretion rate is unchanged, still maintain a 
rather uniform level of turbulence everywhere.

\section{Discussion and conclusions}
We have shown that rms velocities
of $3$--$8$ km s$^{-1}$ in the warm phase of H\,{\sc i}
naturally arise if outer disks are continuously
stirred by a clumpy accretion flow at a rate consistent with observations.
The exact radial variation of the H\,{\sc i} line width depends on the
uncertain contribution of the thermal broadening as a function of
radius and on the adopted accreting mass flux $\dot{\Sigma}(R)$, 
which ultimately depends on which the origin
of the condensing clouds is. 
An interesting possibility is that the outer disk is being rained on
by Galactic fountain material driven by inner starbursts \citep{ben93}
or due to matter circulation in the halo.
Under the rather extreme assumption that clouds fall ballistically 
from infinity,
$\sigma_{W}$ is expected to vary only $\sim 60\%$ between $1$ and $1.7$
times the optical radius. 
The coupling between gas accretion and stellar feedback could account for the 
uniformity of the H\,{\sc i} line widths.

So far, our quasi--isothermal simulations do not capture
all the physics of the multiphase ISM to permit a detailed comparison
between predictions and observations.
Three-dimensional simulations including cooling and heating in thermally
bistable medium, plus a spectrum of mass and radius
of the clouds are being undertaken.  These experiments will
allow us to derive the velocity dispersion for each component (cool, warm 
and hot phases) and compare them with high-resolution H\,{\sc i} observations
that have been able to identify cool and warm neutral components
in the ISM of external galaxies (e.g., \citealt{blo06}).
The accretion of clumpy gas can induce turbulence-compressed 
regions and trigger star formation, which should be also included. 

It is important to consider other phenomena arising in the present accretion
scenario.  For instance,
if turbulence is driven by oblique accretion flows, a large
asymmetry in the scaleheight of the H\,{\sc i} disk as that observed between
the northern and southern halves of the Milky Way \citep{lev06}
is expected because oblique flows have an azimuthal dependence
(see \citealt{san06}). In addition,
infalling clouds moving supersonically at the upper warm galactic disk
develop optical-emitting shocked regions. At a density of the 
preshocked gas of $5\times 10^{-3}$ cm$^{-3}$,
the face-on H$\alpha$ surface brightness $(I_{\alpha})_{\bot}$
of a certain shocked region may range between $10$ and $32.5$ mR
for shock velocities of $50$ and $100$ km s$^{-1}$, respectively
\citep{ray79}. In an edge-on galaxy, the expected surface brightness,
on scales of $\sim 1.5$ kpc,
is the result of the contribution of all the shocked regions through
the sight line: $I\simeq (3\pi/2)N_{cl}R_{cl}^{2}R(I_{\alpha})_{\bot}$,
where $N_{cl}$ is the number of clouds per unit volume, in our
models $N_{cl}\approx 2.5$--$5$ kpc$^{-3}$. At $R=20$ kpc, $I=12$--$40$ mR
depending on the shock velocity, implying
emission measures $EM\equiv \int n_{e}^{2}dl=0.03$--$0.1$
pc cm$^{-6}$, and assuming gas at $8000$ K.
While the detection of such a H$\alpha$ brightness is challenging,
extremely deep H$\alpha$ images of edge-on external galaxies might
constrain the form in which galaxies accrete mass.

\acknowledgements
We thank J.~Cant\'o, A.~Hidalgo-G\'amez, E.~Levine, 
E.~V\'azquez-Semadeni and the referee
for very useful comments. The numerical calculations were performed
in the Computer Center at UNAM. This work 
has been partially supported from DGAPA--UNAM grant IN104306.

\begin{deluxetable}{cccccc}
\tablecaption{Some relevant runs} 
\tablewidth{0pt}
\tablehead{
\colhead{Run$^a$} & $v_{cl}/c_{s}$ & $n_{cl}/n_{0}$ & 
$R_{cl}/h_{0}$ & $10^{2}\lambda_{\rm acc}^{\,\,\,b}$ & \colhead{$\bar{\sigma}_{x}/c_{s}$} }
\startdata
BCa  & $12.5$ & $1$ & $0.23$ & $1.2$  & $0.75$ \\
BCb   & $12.5$ & $1$ & $0.11$ & $6.0$  & $1.25$ \\
BCc   & $12.5$ & $1$ & $0.46$ & $0.24$ & $0.56$ \\
BCd   & $12.5$ & $0.1$ & $0.23$ & $1.2$ & $0.25$ \\
BCe   & $1.25$ & $1$ & $0.23$ & $1.2$ & $0.69$ \\
Sp-a   & $8.33$ & $0.9$ & $0.17$ & $2.2$ & $0.70$ \\ 
Sp-b   & $8.33$ & $0.9$ & $0.17$ & $11.55$ & $1.05$ \\ 
\enddata
\medskip\\

$^a$ BC refers to the extension of the Boucoulares \& Cox (1990) \\
model.  Sp indicates that a Spitzer model with $\alpha=1$ was used. 

$^{b}$ In physical units, an accretion rate of $0.6$ M$_{\odot}$ yr$^{-1}$ 
corresponds \\ 
to $1.2$ in the BC model  and to $2.2$ in
the Spitzer model,  for the \\
Galactic surface density at $20$ kpc.

\label{table:T1}
\end{deluxetable}

\clearpage

\clearpage

\clearpage

\end{document}